\input{aipcheck}

\documentclass[
    ,final            
  ]
  {aipproc}

\layoutstyle{8x11double}

\begin{document}

\title{Multiphase PC/PL relations: comparison between theory and observations}

\classification{<Replace this text with PACS numbers; choose from this list:
                \texttt{http://www.aip..org/pacs/index.html}>}

\author{S. Kanbur}{
  address={SUNY Oswego}
}

\author{M. Marconi}{
  address={Naples Observatory}
}

\author{C. Ngeow}{
  address={University of Illinois}
}

\author{I. Musella}{
  address={Naples Observatory}
}

\author{M. Turner}{
  address={Rice University}
}

\author{S. Magin, J. Halsey, C. Bissel}{
  address={SUNY Oswego}
}

\begin{abstract}
Cepheids are fundamental objects astrophysically in that they hold the key to a CMB independent estimate of
Hubble's constant. A number of researchers have pointed out the possibilities of breaking degeneracies between
${Omega}_{Matter}$ and $H_0$ if there is a CMB independent distance scale accurate to a few percent (Hu 2005).
Current uncertainties in the distance scale are about $10\%$ but future observations, with, for example, the JWST, will
be capable of estimating $H_0$ to within a few percent. A crucial step in this process is the Cepheid PL relation.

Recent evidence has emerged that the PL relation, at least in optical bands, is nonlinear and that  
neglect of such a nonlinearity can lead to errors in estimating $H_0$ of up to 2 percent. Hence it is important to critically
examine this possible nonlinearity both observationally and theoretically. Existing PC/PL relations rely exclusively on
evaluating these relations at mean light. However, since such relations are the average of relations at different
phases. Here we report on recent attempts to compare theory and observation in the multiphase PC/PL planes.

We construct state of the art Cepheid pulsations models appropriate for the LMC/Galaxy and compare the
resulting PC/PL relations as a function of phase with observations. For the LMC, the (V-I) period-color relation at minimum
light can have quite a narrow dispersion (0.2-0.3 mags) and thus could be useful in placing constraints on models. At longer periods,
the models predict significantly redder (by about 0.2-0.3 mags) V-I colors. We discuss possible reasons for this and also compare
PL relations at various phases of pulsation and find clear evidence in both theory and observations for a nonlinear PL relation.
\end{abstract}

\maketitle


\section{Introduction}

The Cepheid PL relation is still regarded as one of the best distance indicators and is considered the key
to the extra-galactic distance scale. The HST Key Project (KP, Freedman et al 2001) used this relation, together
with HST observations of Cepheids in a number of spiral galaxies to estimate $H_0$ to an accuracy of $10\%$.
CMB estimates of cosmological parameters can only estimate the combination $H_0^2{\Omega}$. In order to estimate 
${\Omega}$ with increased accuracy, it is necessary to have a CMB independent estimate of $H_0$ accurate to $1\%$ (Hu 2005).
Motivated by this, there have been recent efforts to improve the accuracy of the Cepheid based estimate of $H_0$ (Macri et al 2006,
Riess et al 2009).

However, recent theoretical (Caputo et al 2000, Kanbur, Ngeow and Buchler 2004) and observational evidence has emerged that the Cepheid PL relation
is not linear (Kanbur and Ngeow 2004, Ngeow et al 2005, Ngeow et al 2009).
The effect of neglecting this possible non-linearity in Cepheid based estimates of $H_0$ can be as much as $1-2\%$ (Ngeow and Kanbur 2006): an effect
that may
be important in the light of efforts to push down errors on $H_0$ estimates to a few percent. Hence it is important to study this possible
non-linearity, both observationally and theoretically.

The Cepheid PL relation is usually both studied and applied at mean light: the mean over all pulsation phases. Yet modern data and its
excellent phase coverage, particularly in the Magellanic Clouds, demand a more detailed approach. Hence we study Cepheid
period-color (PC) and PL relations as a function of phase. Mean light PC/PL relations are clearly the average of the corresponding
relations as a function of phase. Studying these relations as a function of phase can shed light about phases at which possible
non-linearities occur, can lead to important constraints on theoretical models and provide more detailed insights into
pulsation physics.

\section{Methods}

Observationally, we use the excellent data for the LMC provided by the OGLE II/III projects (Udalski et al 1999, Soszynski et al 2008).
For the Galaxy, we use data described in Kanbur and Ngeow (2004). For all these data, we perform a Fourier fit to the data and then re-phase
such that for each star maximum light occurs at phase 0. The we can plot PC/PL relations at any required phase from 0-1.
The models use the code originally developed by Stellingwerf (1982) and adapted to
Cepheid pulsators by Bono, Marconi \& Stellingwerf (1999). The mass-luminosity (ML) relation used is given in Bono et al (2000).
For a given composition, mass and
hence luminosity we compute full amplitude pulsation models at a range of temperatures and convert
the bolometric light curves into magnitude and color variations using static stellar atmospheres
(Castelli, Gratton \& Kurucz 1997a,b). This results in theoretical multiphase PC/PL relations.
Finally we compare both models and theory in terms of PC/PL relations.

\section{Results}
Figures 1-2 represent plots of PL relations as a function of phase using OGLE III data corrected for reddening using the
reddening maps of Zaritsky et al (2004). We see clearly that the PL relation as a function phase has a dynamic nature with
both its linearity and dispersion varying as a function of phase and period. The phase of 0.75 clearly demonstrates a distinct
nonlinearity which is in the form of two lines as opposed to a parabola: there is a sharp break at a period of 10 days.
This agrees with previous work on OGLE II data (Ngeow and Kanbur
2006). Moreover, Ngeow and Kanbur (2006) give arguments indicating why reddening errors cannot be responsible for making an intrinsically
linear relation appear nonlinear. We also note that the PL relation at phases close to 
maximum/minimum light is linear/nonlinear respectively and that the dispersion of the PL relations varies as a function
of phase, with the minimum dispersion occurring at phases close to minimum light (ie around phases 0.7-0.9).

We note that the long period ($\log P > 1$) PC relation is flat agreeing with observations and theory
regarding the Galactic PC relation at maximum light (Simon, Kanbur and Mihalas 1993, Kanbur, Ngeow and Buchler 2004).

Figures 3-4 represent a comparison with models for
Galactic data and figures 5-8 are a comparison with models using OGLE II LMC data.In all these figures, the crosses are
observations, corrected for
reddening using published reddening values in the literature or derived by the OGLE team (Udalski et al 1999). The various symbols 
portray Cepheids of different masses, usually ranging from 5, 7, 9 and 11 solar masses. The adopted ML relation
then determines the luminosity once the mass and chemical composition are chosen. For each mass, we compute models with different temperatures, with the
coolest models having the longest period. For Galactic data, we use a solar type composition ($Z=0.02, Y=0.28$) 
and a composition ($Z=0.01$) for LMC data. Figures 2-6 are PC relation plots using the $(V-I)$ color and figures 7-8
are PL relation plots is in the $V$
band.

A comparison of the Galactic and LMC PC relations at maximum/minimum light clearly suggests that Galactic Cepheids with $0.4 < \log P < 0.8$
obey a different relation to longer period Cepheids - something that is not true for LMC Cepheids. 
Further the Galactic PC relations at maximum light are flat from about $\log P > 0.8$ onwards whereas the LMC maximum light
relation is flat for Cepheids with $\log P > 1$.
The observed PC relations in figures 5-6 using OGLE II data are broadly similar to those in figure 1 using OGLE III data.
The dispersion of the PC/PL relation changes not only with phase, where the tightest dispersion is close to minimum light, but also
with period, where the tightest dispersion appears close to $\log P \approx 1$. 

An important result is that changes in the PC relations as a function of phase are mirrored by changes in the PL relation
as a function of phase. This is to be expected since the PC/PL relations are both different aspects of the Period-Luminosity-Color
(PLC) relation.

The comparison between models and theory in figures 3-4 for Galactic PC relations appears to rule out 11 solar masses as a 
possible mass for longer period Cepheids, or at least suggest that the instability strip at such
high masses is significantly bluer than predicted. However this could be influenced by a possible lack of data of long period Cepheids.
For LMC Cepheids, 11 solar masses can match the observations for a certain range of temperatures. The 7 solar mass model seems
to capture the observed behavior of LMC Cepheids in both PC/PL planes. This particular sequence shows a multi-valued trend with
temperature which reproduces the change in slope seen in both PC/PL plots at phases around minimum light.
We note that this observation may have been harder to spot if the comparison had been done in the mean light PC/PL plane
and that a comparison in the multiphase PC/PL plane can, in principle, place very stringent constraints on models.


\begin{figure}
  \includegraphics[height=.3\textheight,angle=-90]{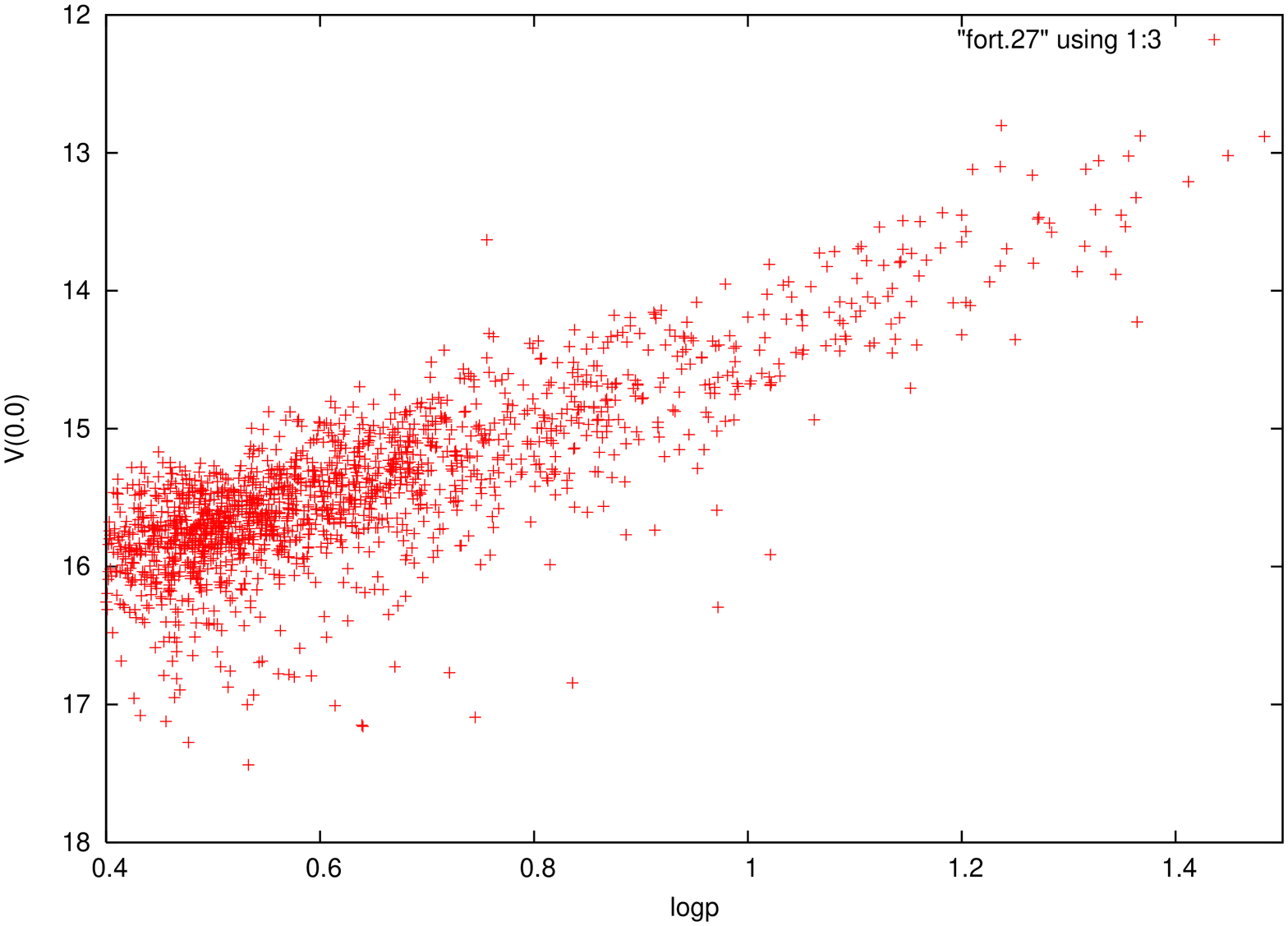}
\hfill
  \includegraphics[height=.3\textheight,angle=-90]{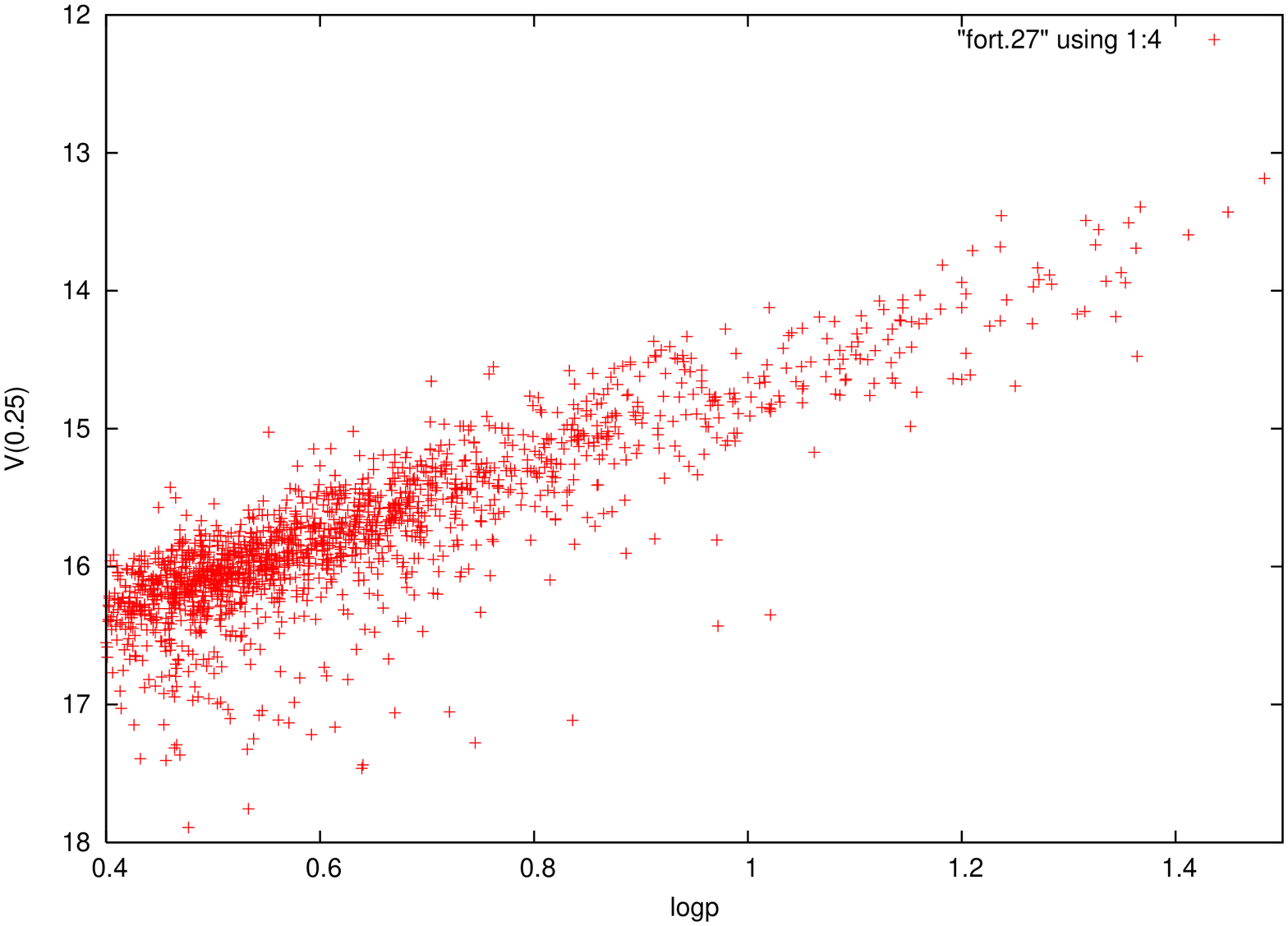}
\hfill
 \caption{OGLE III PL relations at phase 0 (left panel) and phase 0.25 (right panel).}
\end{figure}

\begin{figure}
  \includegraphics[height=.3\textheight,angle=-90]{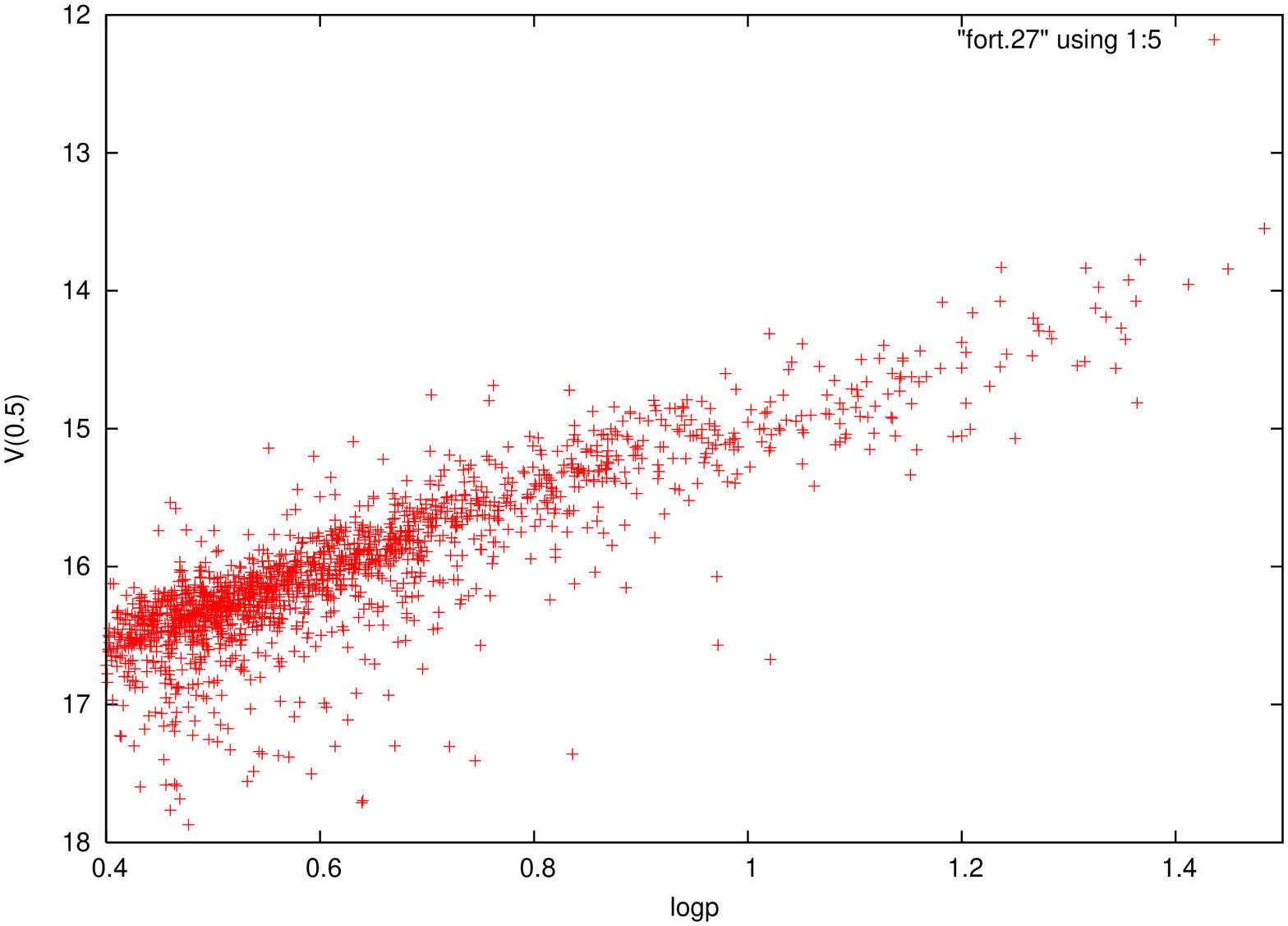}
\hfill
  \includegraphics[height=.3\textheight,angle=-90]{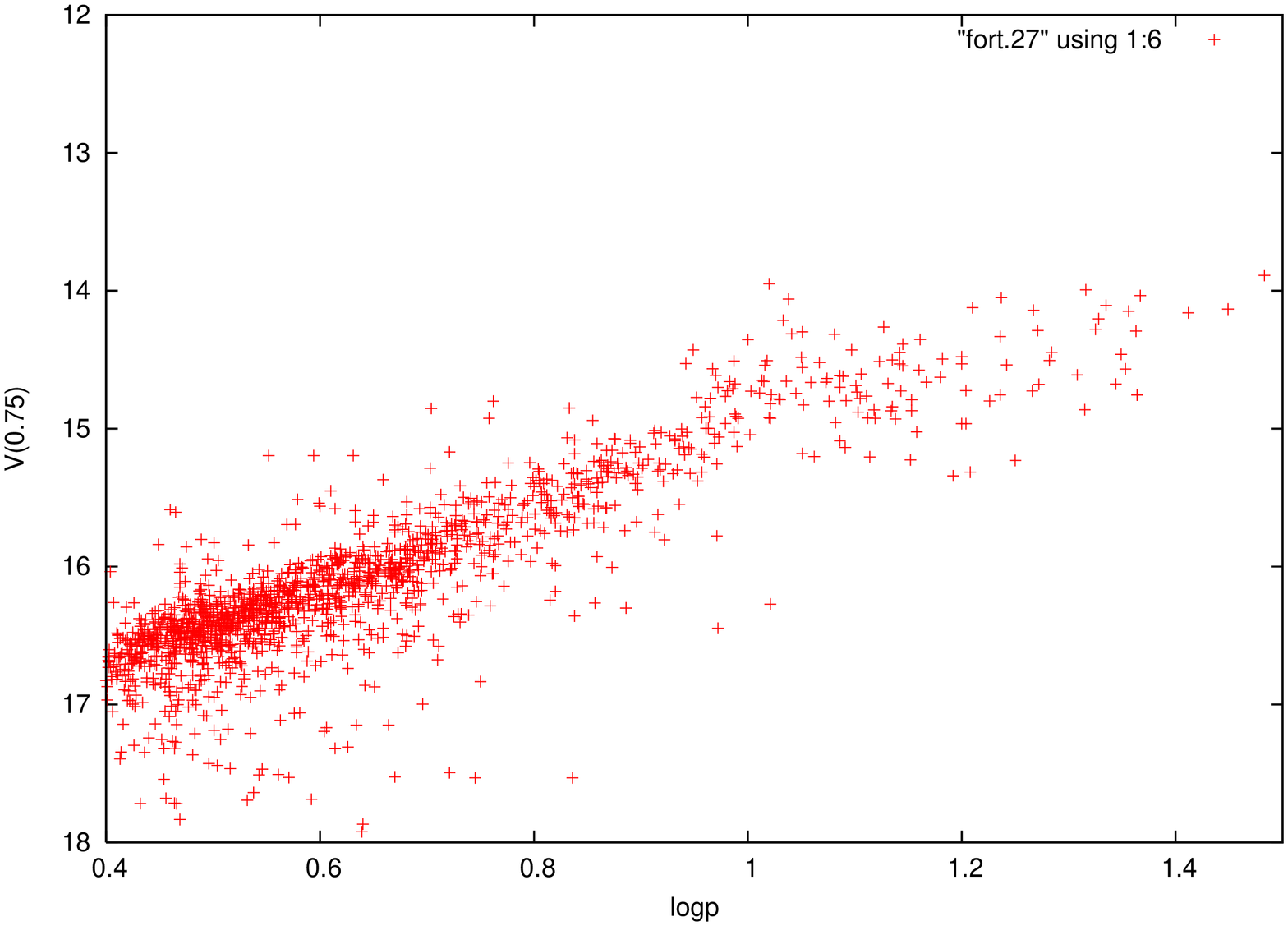}
  \caption{OGLEIII PL relations at phase 0.5 (left panel) and phase 0.75 (right panel).}
\end{figure}

\begin{figure}
  \includegraphics[height=.20\textheight]{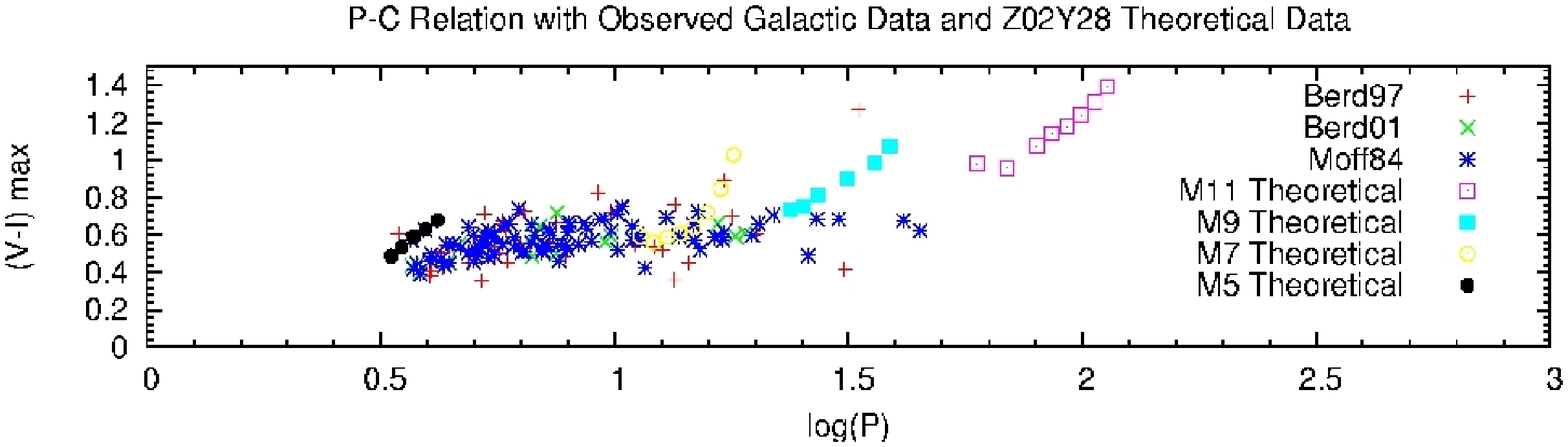}
\caption{Galactic PC relation at maximum light overlaid with model predictions (see text).}
\end{figure}

\begin{figure}
\includegraphics[height=.20\textheight]{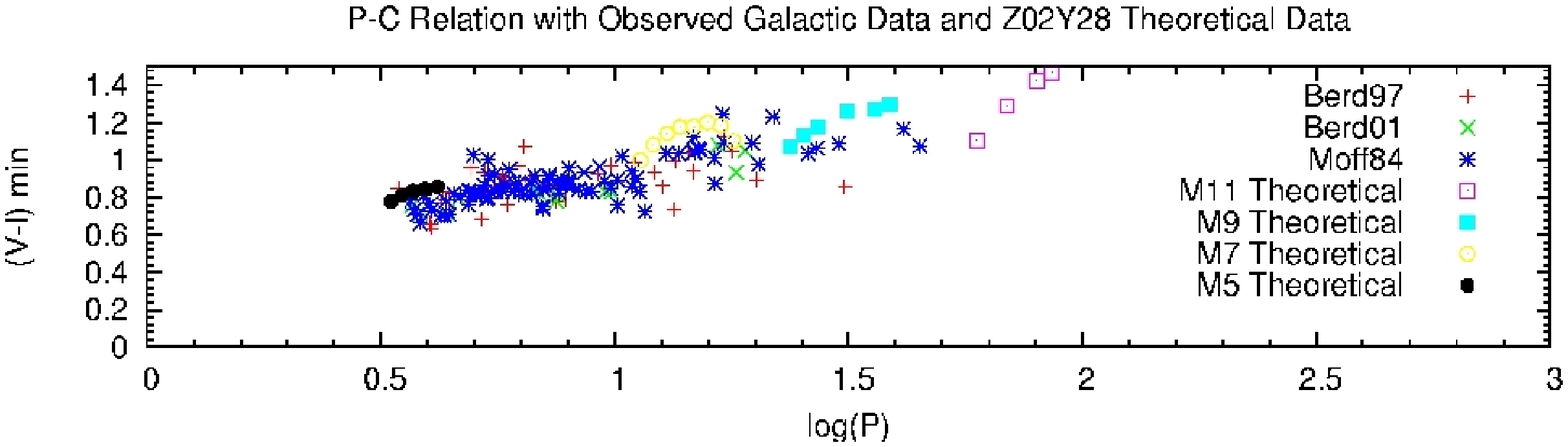}
  \caption{Galactic PC relations at minimum light overlaid with model predictions (see text).}
\end{figure}

\begin{figure}
\includegraphics[height=.2\textheight]{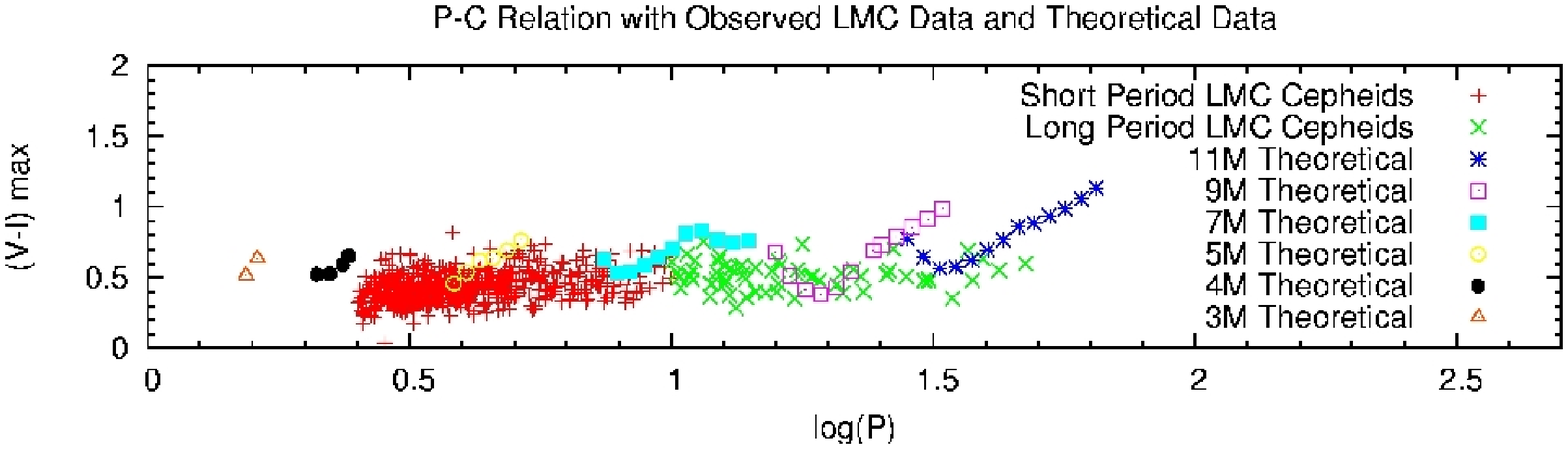}
\caption{LMC PC relations at maximum light overlaid with model predictions (see text).}
\end{figure}

\begin{figure}
\includegraphics[height=.2\textheight]{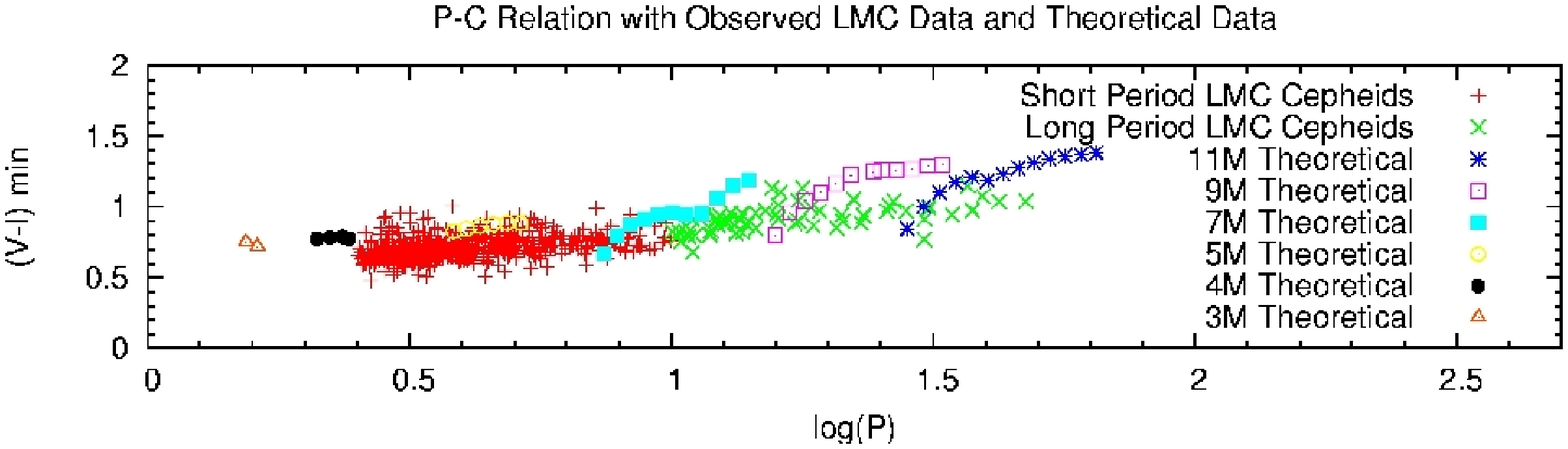}
\caption{LMC PC relations at minimum light overlaid with model predictions (see text).}
\end{figure}

\begin{figure}
\includegraphics[height=.2\textheight]{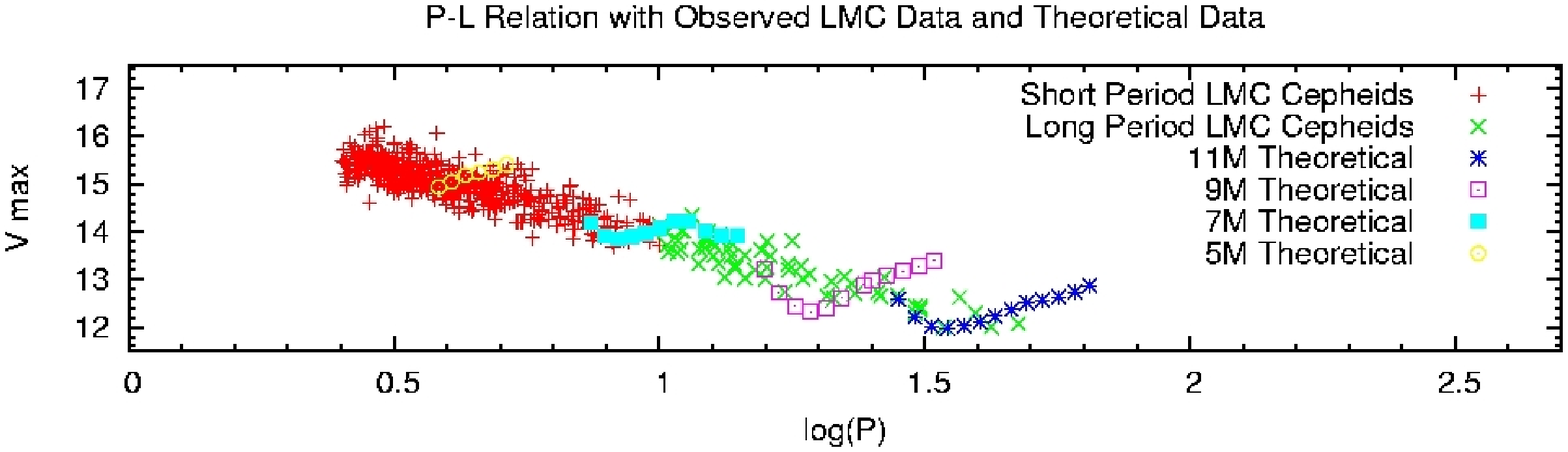}
\caption{LMC PL relations at maximum light overlaid with model predictions (see text).}
\end{figure}

\begin{figure}
\includegraphics[height=.2\textheight]{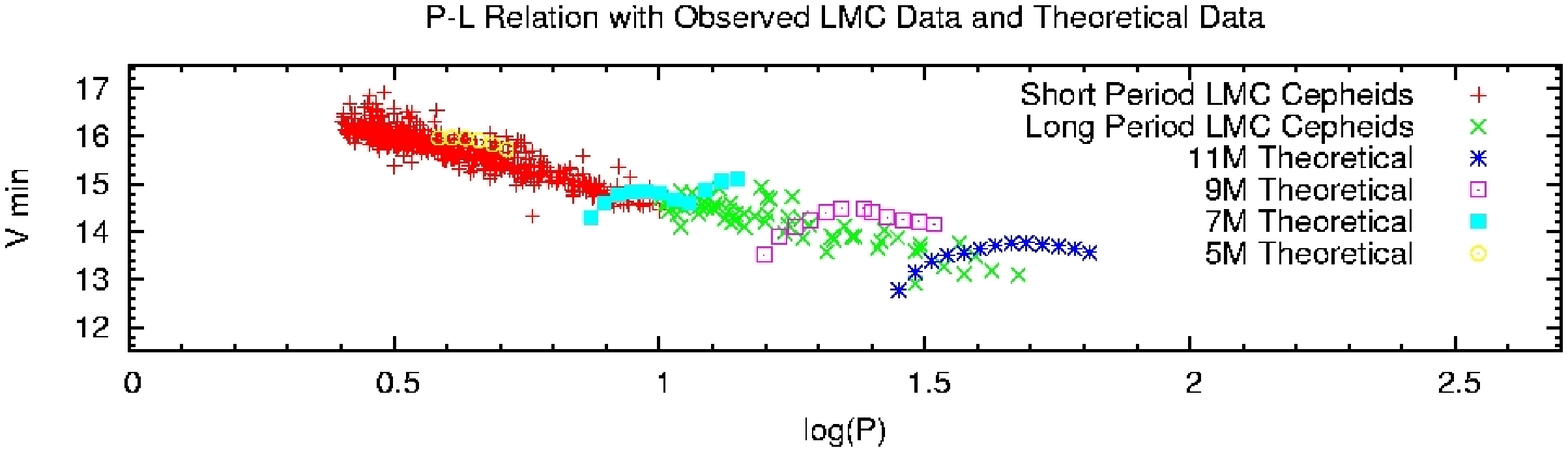}
\caption{LMC PL relations at minimum light overlaid with model predictions (see text).}
\end{figure}


\begin{theacknowledgments}
  SMK acknowledges support from the Chretien Award of the American Astronomical Society and
SUNY Oswego.
\end{theacknowledgments}



\bibliographystyle{aipproc}   





\end{document}